\documentclass[12pt]{article}
\usepackage{a4}
\usepackage{epsfig}
\usepackage{amssymb}
\begin{document}

\setlength{\parskip}{1.5ex}
\thispagestyle{empty}

\begin{flushleft}
   DESY 00-018 \\ 
   hep-ex/0002002 \\
   February 2000
\end{flushleft}

\vspace*{0.5cm}

\begin{center}
 {\LARGE\bf Observation of coasting beam at the\\[1.5ex]
 HERA Proton--Ring}\\[4ex]
{\Large \bf HERA--B Target Group}\\[2ex]
{\large \bf K. Ehret, M. Funcke, S. Issever, T. Jagla$^+$, S.
  Schaller$^{++}$, S. Spratte, M. Symalla, Y. Vassiliev, D. Wegener \\Institute of
  Physics\\University of Dortmund}\\
\vspace*{0.5cm}
{\large \bf T. Lohse\\ Institute of Physics\\ Humboldt University Berlin}

\end{center}

\vspace*{2.3cm}

\underline {\large\bf Abstract:}\\

We present data collected with the HERA--B wire target which prove the
existence of coasting beam at the HERA proton storage ring. The
coasting beam is inherently produced by the proton machine operation
and is not dominated by target effects.

\vspace*{2.0cm}

\begin{tabular}{ll}
$^+$&Now at Max--Planck--Institut für Kernphysik, Heidelberg, Germany\\
$^{++}$&Max--Planck--Institut für Kernphysik, Heidelberg, \\
& now at Max--Planck--Institut für Physik, Werner Heisenberg Institut,\\
& München, Germany
\end{tabular}

\newpage

\section{Introduction}
The internal target of the HERA--B experiment consists of eight ribbons
positioned around the proton beam of the HERA storage ring at a distance of 
$3 - 6~rms$ beam widths \cite{Eh1}. Protons drifting away from the beam core
interact with these targets and are expected to produce among other hadrons 
B--mesons. Since 1992 we studied those problems of basic importance for the 
experiment \cite{Ha, Lo} namely the interaction rate achievable and its 
fluctuations, the target efficiency, the spatial distribution of the 
interaction vertices, and the interference with the beams as well as the 
background induced by the wire target at the place of other experiments 
running in parallel to HERA--B. It turns out that the design goal of 
a target efficiency $\ge 50 \%$ and the interaction rate of $\sim$ 40~MHz 
can be achieved \cite{Eh2}.

The HERA--B detector with its readout and different trigger levels relies 
on the close correlation between the time of the interaction and the bunch 
crossing signal of the proton beam. Therefore this correlation was studied 
in some detail. Besides the interactions due to the bunched protons an 
unexpected high background from bunch uncorrelated interactions was 
observed, which showed a strong asymmetry in the transverse plane. In this 
paper we summarize the observations, quantify them and discuss possible 
sources of this background.

\section{Coasting Beam}

\begin{figure}[ht!]
\begin{center}
\epsfig{file=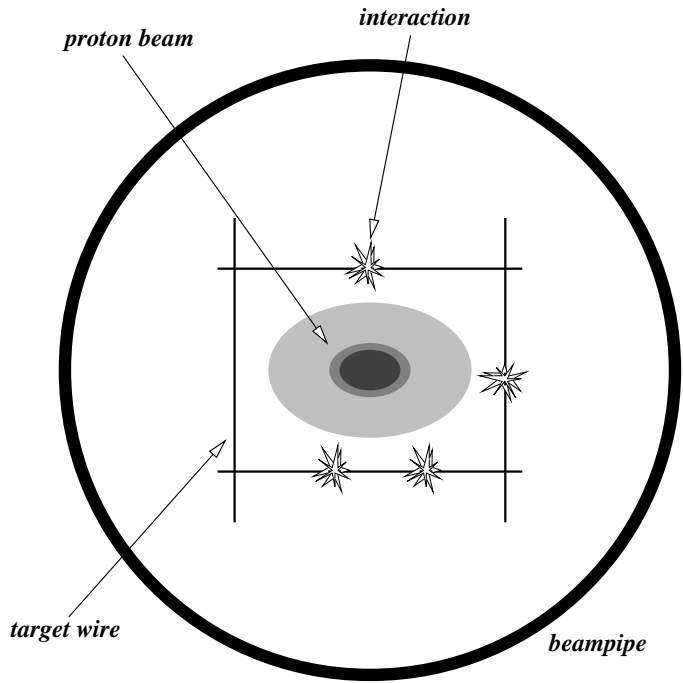, height=9cm}
\end{center}
\caption{Schematic view of the HERA--B wire target.}
\label{abb1}
\end{figure}

\begin{figure}[ht!]
\begin{center}
\vspace*{1cm}
\epsfig{file=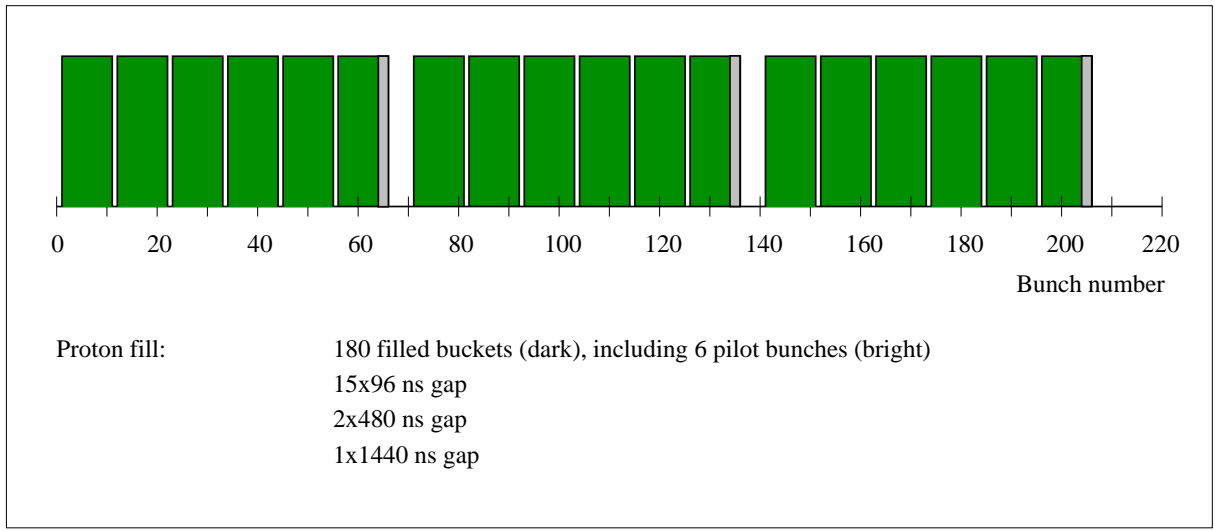, height=7cm}
\end{center}
\caption{Schematic representation of the bunch structure of a HERA
  proton--ring fill.}
\label{abb2}
\end{figure}

A schematic view of the target is shown in fig.\ref{abb1}. Protons in the
halo or close to the beam core interact with the target
wires which are positioned inside/outside (``inner'' and 
``outer'' wire) and up/down (``upper'' and ``lower''
wire) with respect to
the center of the storage ring and the 
beam respectively. A sketch of the bunch structure of the HERA
proton beam is shown in fig.\ref{abb2}. The bunches are organized in $3
* 6$ trains, each consisting of 10 bunches of $\sim 1~ns$ length. The
bunch spacing within a train corresponds to $96~ns$. The trains are
separated by an empty bunch. 6 trains correspond to one fill of the
PETRA preaccelerator, they are separated from the consecutive 6 trains
by a gap of $480~ns$.
The last 15 buckets are empty to enable a safe beam dump (kicker gap).
A complete revolution of a fill corresponds to 
$220 * 96~ ns = 21.12~\mu s$.

\begin{figure}[ht!] 
\begin{center}
\epsfig{file=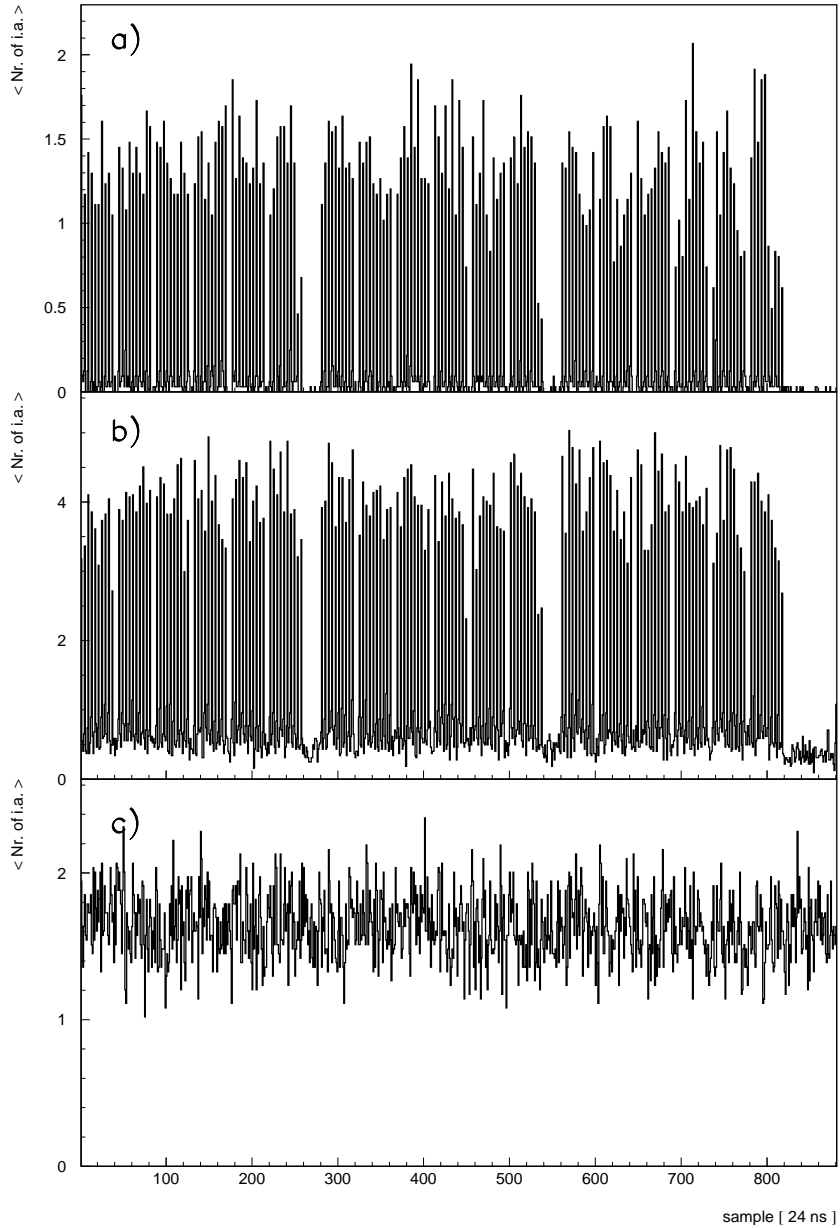, height=17cm}
\end{center}
\caption{Three typical results observed for the time structure of the
  proton interaction with the wire target are shown a) inner wire
  at $\sim 4~\sigma$ from the beam center, b) outer wire $\sim
  4~\sigma$ from the beam center and c) outer wire $\sim
  6~\sigma$ from the beam center. Note the measured interaction rate
  in the region of empty buckets especially the kicker gap for sample
  numbers $>~820$.}
\label{abb3}
\end{figure}

\begin{figure}[ht!]
\begin{center}
\epsfig{file=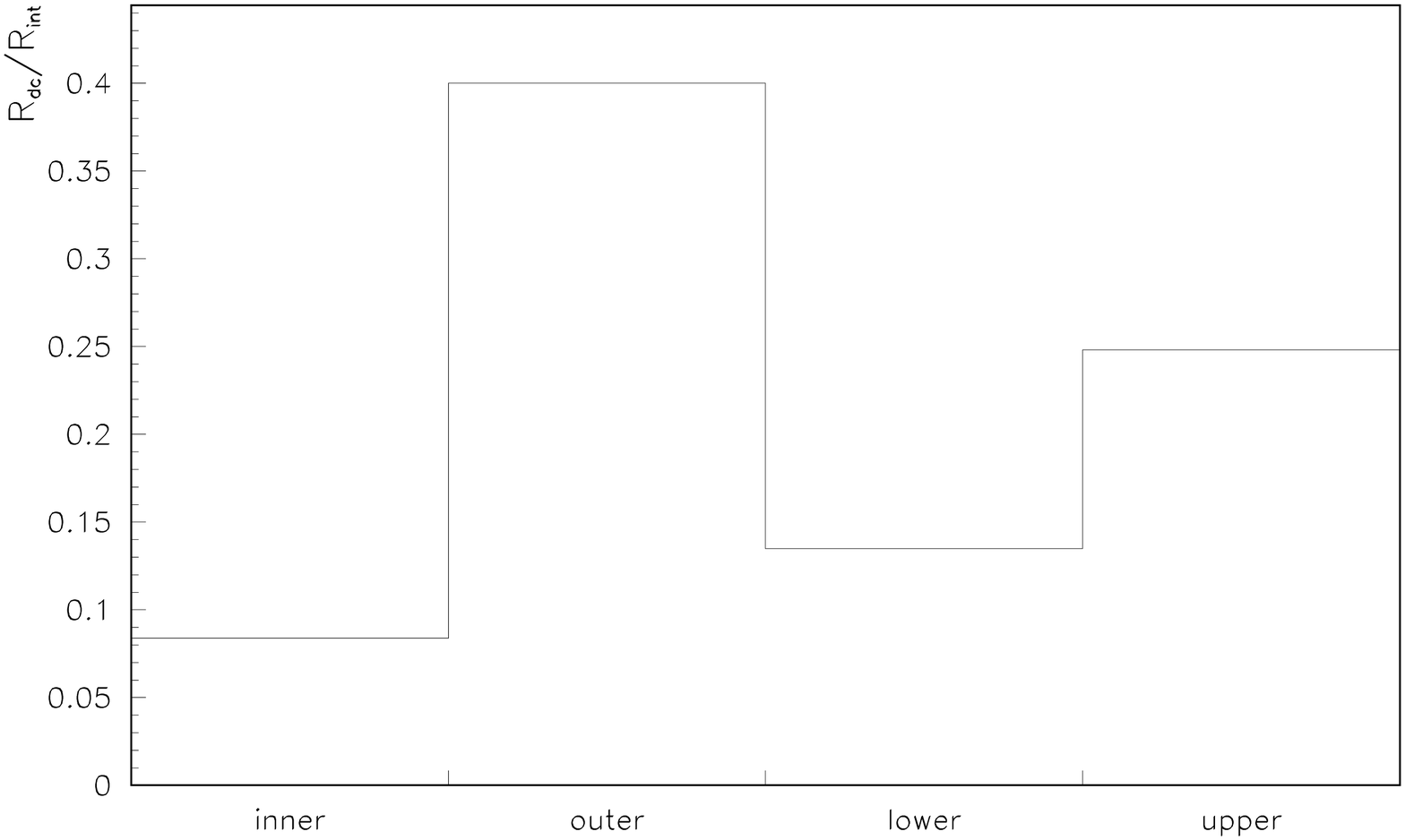, width=13cm}
\end{center}
\caption{Relative contribution of unbunched protons to the total
  interaction rate for an inner, outer, lower and upper
  wire respectively. The data were collected 1999 at a typical total
  interaction rate of $\sim$ 10~MHz.}
\label{abb5}
\end{figure}

A FADC system is used to record separately the contribution of each single
bunch to the interactions produced at the target measured with scintillator 
hodoscopes \cite{Sc}. The FADC samples with the fourfold rate of the bunch 
crossing, i.e. every $24~ns$ a signal is recorded.
Per readout cycle of the FADC--system 880 bytes are recorded allowing to
study the interactions in a time slice of 21.12 $\mu s$ which
corresponds to the time needed by a proton to cycle once around the
HERA ring. A rate of 150 Hz for the readout is achieved. About 5000
succeeding measurements are summed up to accomplish a reasonable statistics,
hence the FADC rates presented in this paper are averaged over a time
of $\sim 30~s$. As demonstrated by fig.\ref{abb3}a  the target--beam halo
interaction indeed shows the bunch structure of the proton beam if the
protons interact with an inner target wire. The time structure of
the outer target wires  (fig.\ref{abb3}b,c) differs qualitatively from the
inner one, though the data were collected consecutively within a short time 
interval. While the inner wire provides a clear bunched signal, for the 
outer wire in addition to the bunch correlated events a continuous 
background is observed (fig.\ref{abb3}b). Even in the regions of the empty 
RF--buckets and of the kicker gap a strong signal shows up if the outer
wire is positioned in the beam halo. If the outer--target is hit by
protons at a distance of 
${\raisebox{0.5ex}{$\scriptstyle\gtrsim$}}~ 5~\sigma$ 
from the beam center only the $dc$ component of the beam contributes 
(fig.\ref{abb3}c). We have convinced ourselves that the observed 
interactions from the continuous component have the same signature as the 
bunch correlated events: exploiting the HERA--B Si--vertex detector it can 
be shown that the bunched and the $dc$ component of the proton beam interact
with the wire, moreover the relative rates of different detector components 
for the two types of events are equal.
The difference of the time distributions for protons interacting with the
inner and outer wire respectively hint to a $dc$ component of the machine 
current with an energy smaller than the synchronous protons 
$(\Delta E/ E<0)$  since at the  position of the target the horizontal 
dispersion $D_x$ of the HERA proton--ring is negative and therefore the 
offset $\triangle x$ of the beam due to its energy deviation 
$\frac{\Delta E}{E}$ is positive, $\triangle x = D_x \frac{\Delta E}{E} > 0$.
The ratio of the integrated continuous background to the total interaction 
rate is smallest for the inner wire, followed by the lower and upper wire 
and it is largest for the outer wire (fig.\ref{abb5}).

\begin{figure}[ht!]
\begin{center}
\epsfig{file=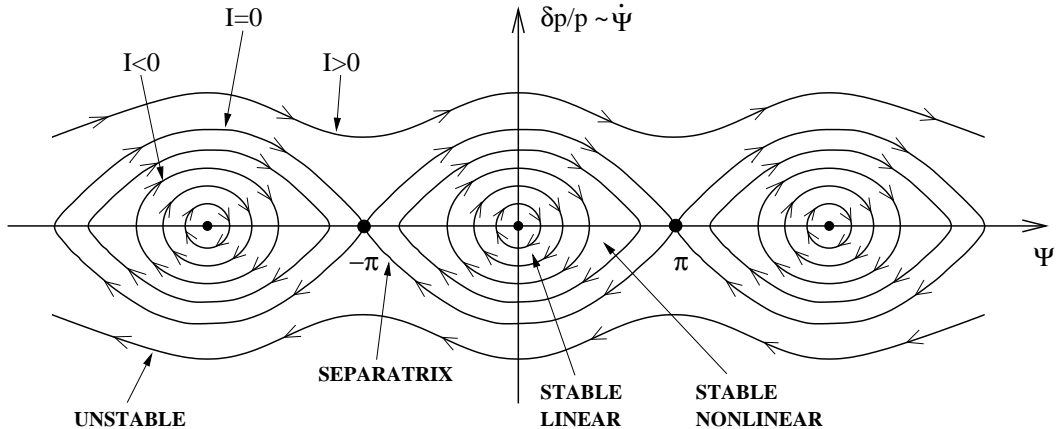, width=14cm}
\end{center}
\caption{Simplified sketch of longitudinal phase space of the HERA proton 
 storage ring.}\label{abb51}
\end{figure}

These observations can be explained easiest by discussing the particle
trajectories in longitudinal phase space (fig.\ref{abb51})
\footnote{The HERA proton ring is operated with cavities excited at
  two different RF frequencies (52~MHz and 208~MHz). The figure is a
  simplified sketch for the case of a single frequency.}, where
$\Psi$ describes the longitudinal phase of a particle in a bunch
and $\dot \Psi \sim \frac{\delta p}{p}$ the longitudinal momentum deviation
of a particle with respect to the centroid. The boundary
(separatrix) between the region of stable and unstable motion is
characterized by the invariant

$$I \approx \left( \frac{\delta p}{p} \right) ^2 - \left( \frac{2
    Q_s}{h~\alpha_p}~ \cos ~\frac{\Psi}{2} \right)^2 = 0$$

\begin{tabular}{lll}
where & $Q_s$ & synchrotron tune\\
& $h$ & harmonic number\\
& $\alpha_p$ & = $\frac{\delta L/L}{\delta p/p}$ momentum compaction
factor\\
& $L$ & path length of proton for 1 turn
\end{tabular}

Particles in the stable region $(I<0)$ stay bunched while particles
outside the stable region $(I>0)$ are debunched. These unbunched protons
in case of the HERA storage ring deviate, depending on the voltage of
the RF--system,  by 
$\frac{\Delta E}{E} > (2 \dots 3) \cdot 10^{-4}$ from the
centroid particle. Note that the dynamic energy acceptance of the
machine $\frac{\Delta E}{E} \approx 10^{-3}$ is much larger than
$\left( \frac{\Delta E}{E} \right)_S = (2 \dots 3) \cdot 10^{-4}$, the 
energy deviation of typical protons close to the separatrix. The
results shown in figs.\ref{abb3}, \ref{abb5} demonstrate the evidence 
of an unbunched 
beam component with $\Delta E < 0$, the so called coasting beam.

Since these protons have smaller energies than the bunched particles an 
energy loss mechanism has to exist which forces the bunched protons to pass 
the separatrix and debunch. Besides machine inherent sources as synchrotron 
radiation (\mbox{$\sim 10~ eV$} per turn), noise of the RF--system and
intrabeam scattering the energy loss of protons in the target can produce the 
coasting beam. As far as these losses can be quantified they are much smaller
than the maximum energy deviation of $\Delta E = 0.27~ GeV$ allowed for the 
stable longitudinal phase space at standard RF--voltage. Hence these energy 
losses are expected only to force protons near the separatrix to pass the
phase space boundary of stable bunched beams.

\subsection{Source of coasting beam}

\begin{figure}[ht!]
\begin{center}
\epsfig{file=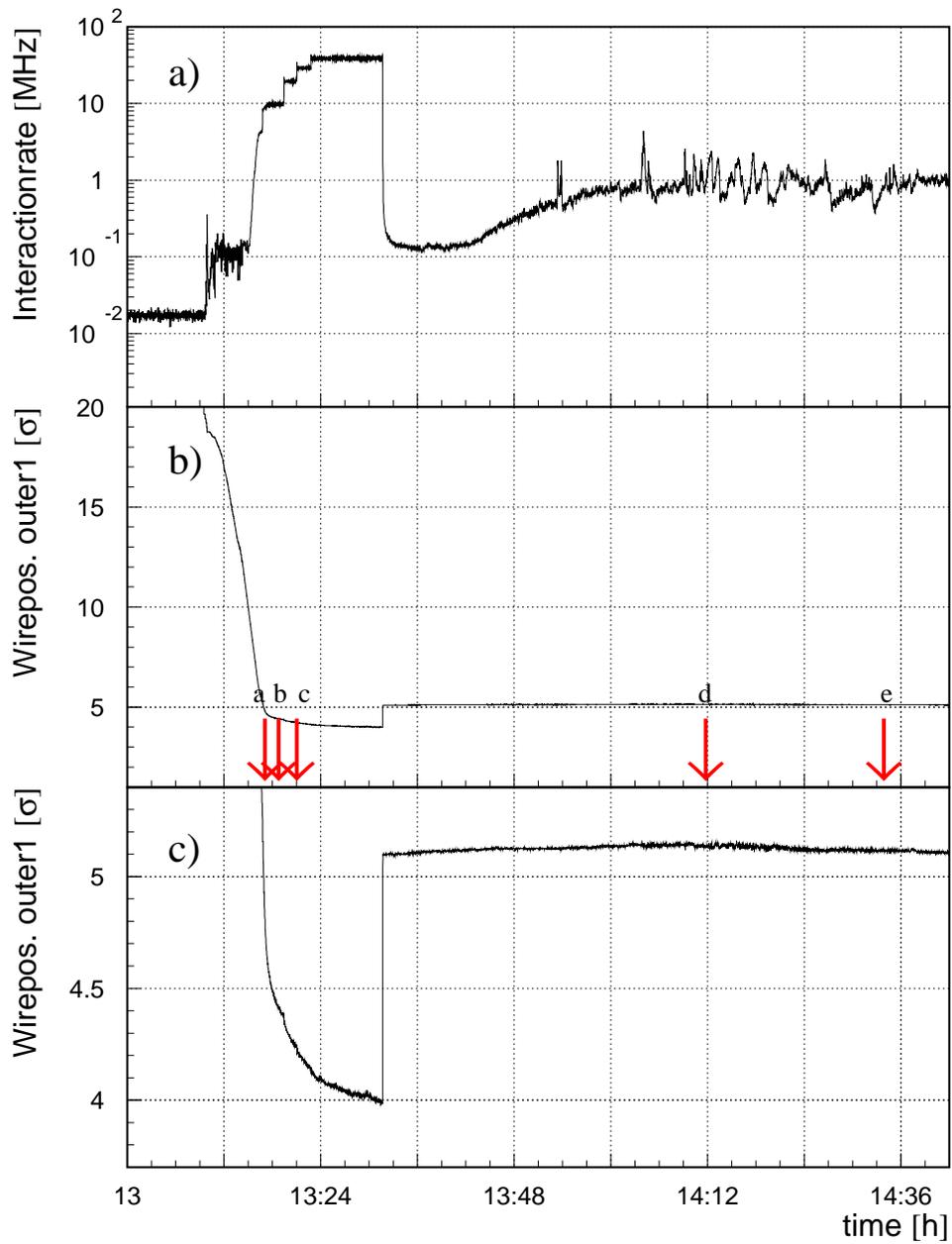, width=13.5cm}
\end{center}
\caption{ a) Measured total rate, b)  wire positions as function of
  time for a virgin beam, c) same as b) with expanded vertical scale.
  The measurements were performed with an outer wire.}\label{abb6}
\end{figure}

\begin{figure}[ht!]
\begin{center}
\epsfig{file=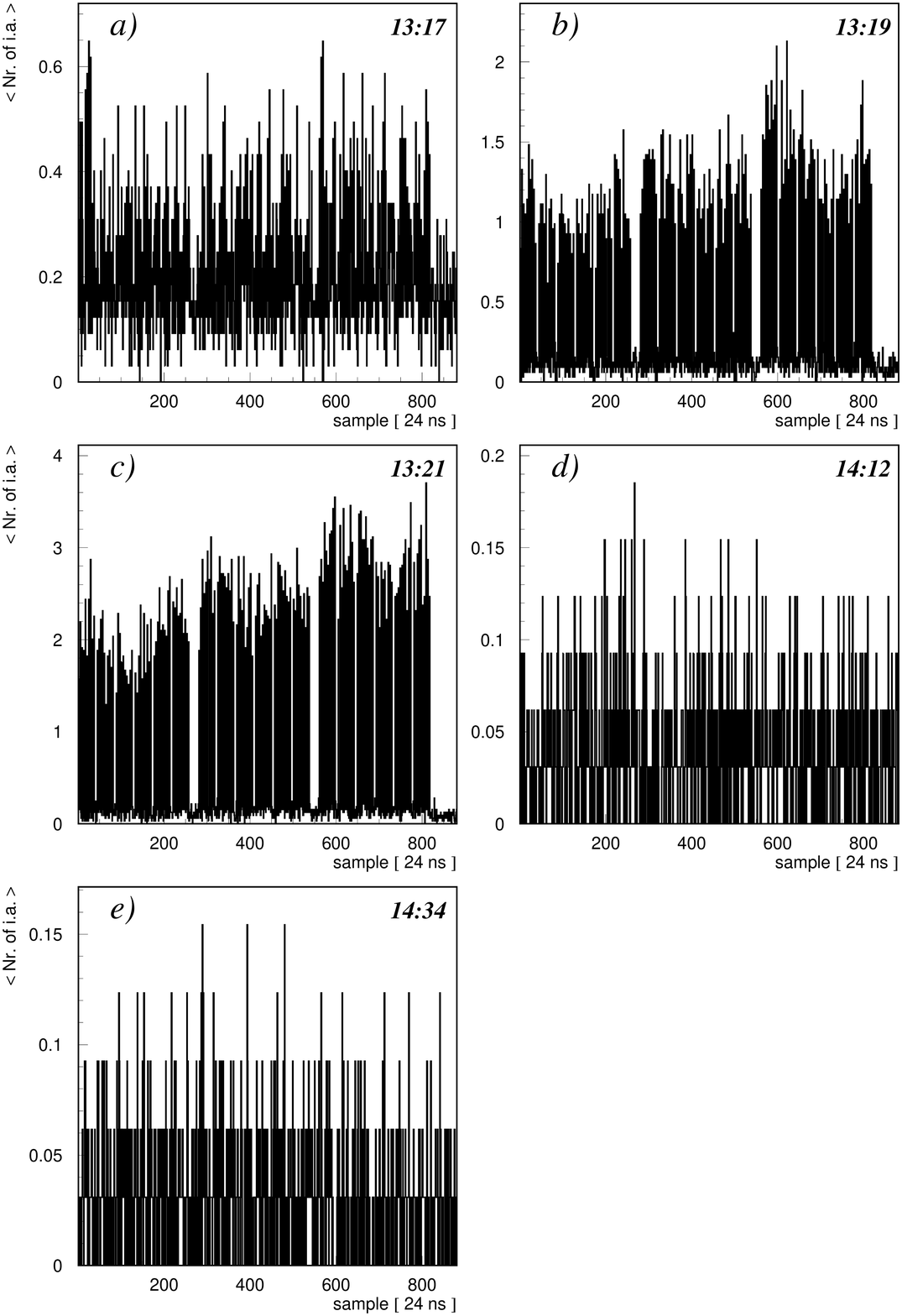, width=12.5cm}
\end{center}
\caption{Rate distribution  measured at
 different times with a FADC. Only the proton ring was filled during
 the measurements. The plots a) -- e) correspond to the consecutive
 times indicated by the arrows in fig.\ref{abb6}.}\label{abb7}
\end{figure}

The first study was performed with a virgin beam, no electrons were stored 
in the electron--ring of the HERA storage--ring complex and the measurement 
started a few minutes after stable beam conditions were declared. In 
fig.\ref{abb6}\,b,c the position of the ``outer'' target with respect to the
beam  center of gravity is shown, the measured total interaction rate is 
plotted in fig.\ref{abb6}a. At different times, indicated by the arrows on 
the time axis, FADC spectra  were analyzed which allow to separate the bunch
contribution from the coasting beam. In fig.\ref{abb7} the FADC
spectra are plotted for the different time intervals. The data shown
in fig.\ref{abb7}a were collected when the target wire touched 
the beam for the first time. In this case the continuous coasting beam 
contribution to the interactions dominates.
Since before this measurement no target wire had touched the
beam this measurement unquestionably proves that the machine itself
produces a coasting beam. Approaching further
the beam core the bunch contribution starts to develop. 

In the next step the target was withdrawn from the beam core and placed 
at a fixed position 
(fig.\ref{abb6}\,b,c). As shown by fig.\ref{abb6}a the total interaction 
rate is strongly 
reduced but after a short time starts to increase again.
Figs.\ref{abb7}d,e demonstrate
that  the coasting beam ``diffuses'' faster towards the outer
target than the bunched component, since no bunch correlated contribution 
is detected.

\begin{figure}[ht!]
\begin{center}
\epsfig{file=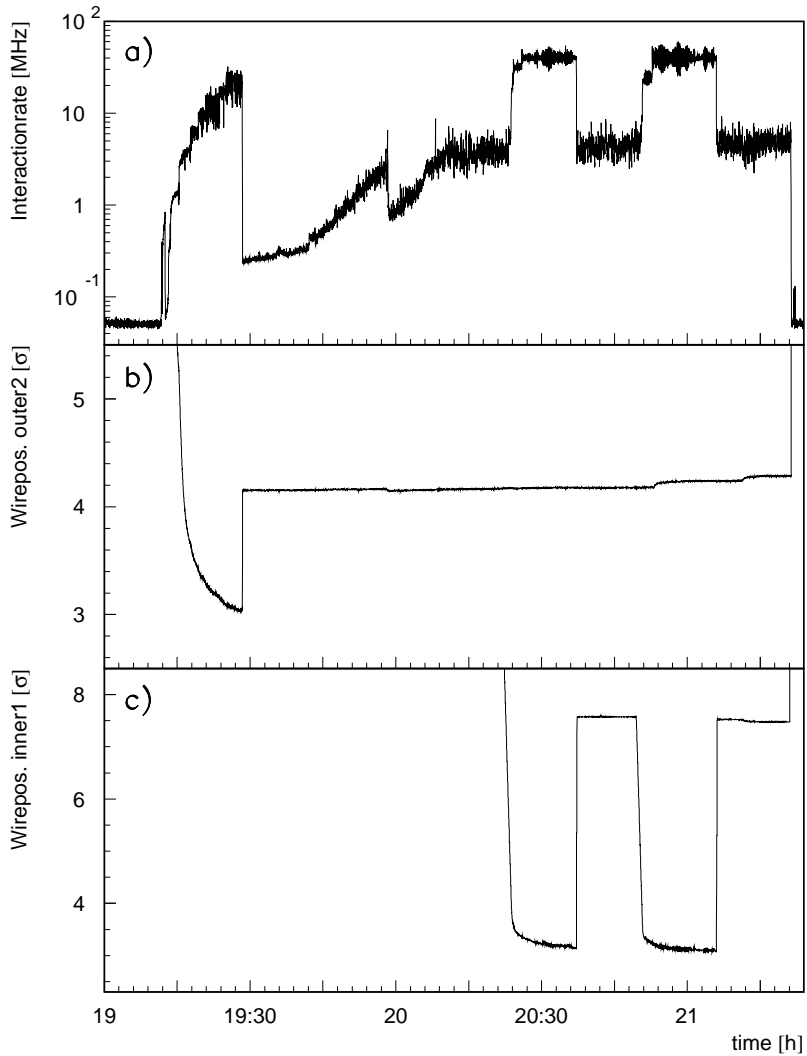, height=15.5cm}
\end{center}
\caption{Total interaction rate (a) and position of outer (b) and
  inner wire (c) as function of time. The glitch observed at $\sim$ 20:00
  in the rates is due to a sudden change of the beam position which
  recovers after $\sim 10~min$.}\label{abb7.1}
\end{figure}

Further evidence for a machine induced coasting beam component follows
from a second measurement. The total interaction rate and the position of 
the wires (inner I and
outer II) at different times are shown in fig.\ref{abb7.1}. At 19:15 the
outer wire is moved towards the beam. An increase of the interaction
rate and its fluctuations is observed. At 19:30 the outer wire is
retracted, the interaction rate decreases strongly and recovers within
the next half hour to a level of several MHz. This behaviour was already 
observed in fig.\ref{abb6}.
Starting at 20:18 the inner wire is moved towards the beam
(fig.\ref{abb7.1}c). As expected, the interaction rate increases. At 20:35 
the inner wire is retracted by $2~mm$ from the beam core, the
interaction rate is strongly decreasing to the level observed for $t <$
20:18. This behaviour is reproducible as shown by the measurements
performed in the time interval 20:42 $\le t \le$ 21:12. 
\vspace*{-.5cm}
These results demonstrate that the energy loss of bunched particles due to
interaction of the protons with the
(inner) target wire is not a major source for producing the coasting beam.
Comparing the interaction rate measured at $t$ = 20:18 and $t$ = 20:48 
$h$, it follows that $\le 15 \%$ of the coasting beam interacting
with the outer target was produced by energy loss of bunched protons
in the inner target.

\begin{figure}[ht!]
\begin{center}
\epsfig{file=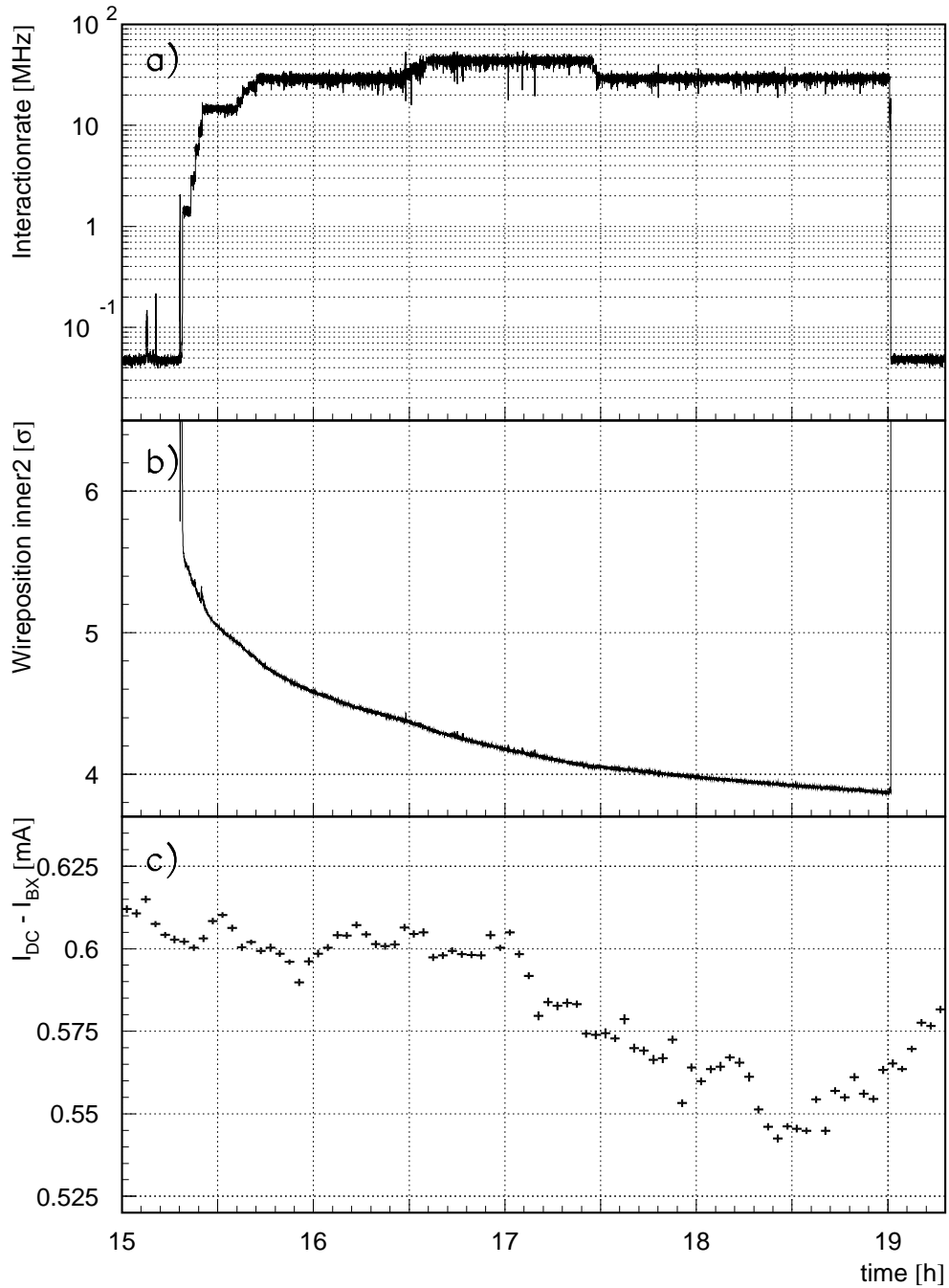, width=14cm}
\end{center}
\caption{Measured total rate (a), position of the inner wire with
  respect to the beam center (b) and $dc$ current as measured with a
  current monitor (c).}\label{abb7.2}
\end{figure}

A similar conclusion can be drawn from fig.\ref{abb7.2}. In this case
the coasting beam component is detected (fig.\ref{abb7.2}c) by a $dc$
current monitor. This current in good approximation stays constant for
15:00 $\le t \le$ 19:00, when the inner target was positioned at
$5.4~ \sigma$ to $3.9~ \sigma$ from the beam core
(fig.\ref{abb7.2}b) and produced a high interaction rate of $>$ 30~MHz
(fig.\ref{abb7.2}a). 
Note, however, that in other measurements a slight increase of
the current is observed for the same setup of the wire target. 
As demonstrated by fig.\ref{abb7.2}c for $t \ge$ 19:00 an increase of the
$dc$ current is detected though the wire is pulled away from the beam
(fig\ref{abb7.2}b). Considering these effects one arrives again at a
target contribution to the coasting beam \mbox{of $<$ 20\,\%.} Moreover, 
these observations stress the sensitivity of the observations on details of
the machine setup, collimator positions etc. 

\begin{figure}[ht!]
\begin{center}
\epsfig{file=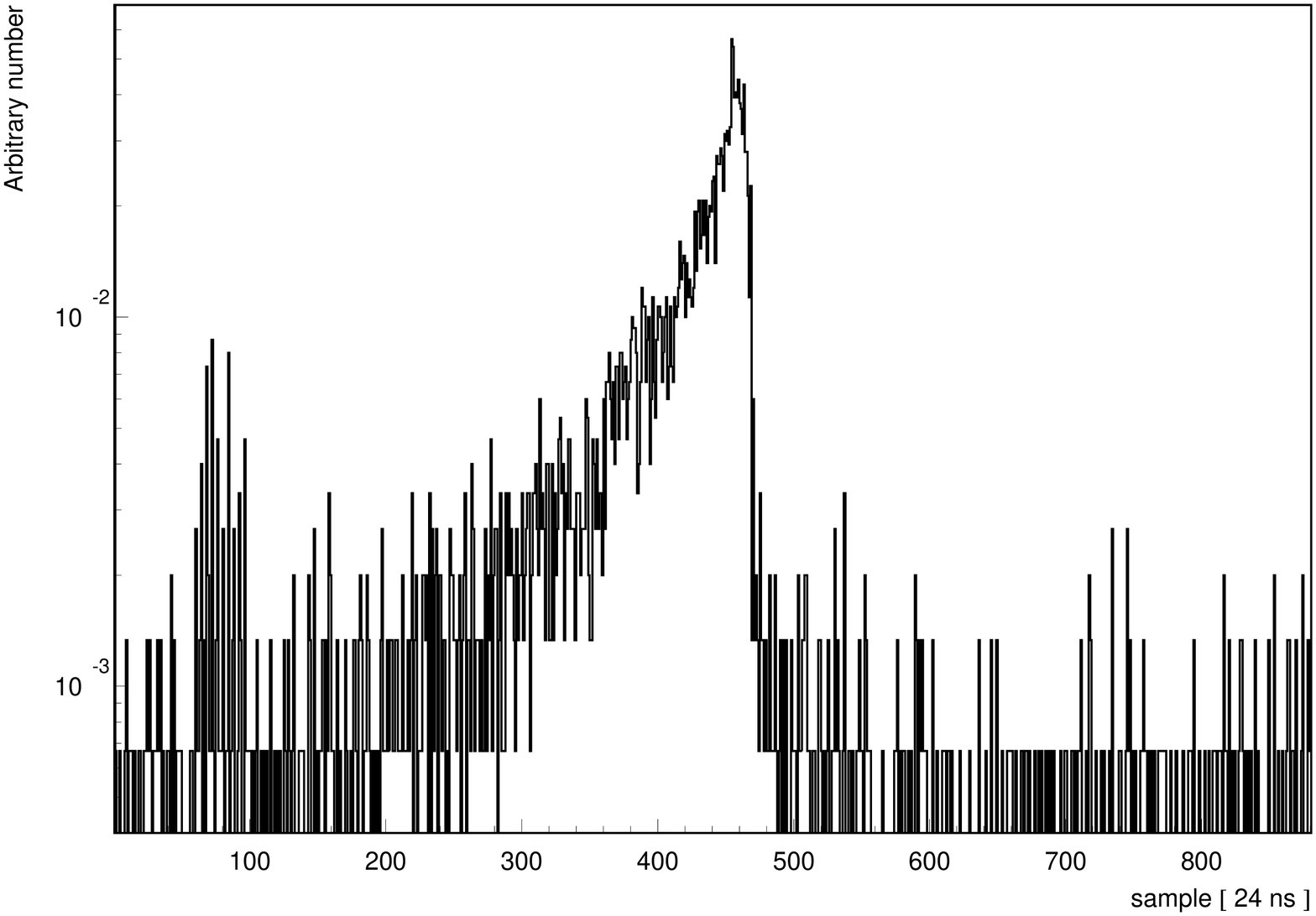, width=12cm}
\end{center}
\caption{FADC spectra of a virgin proton fill where 10 bunches are
  stored in the time interval channel 60 -- 100. The coasting beam is
  excited by a short kicker pulse every 21.12 $\mu s$ at a time
  corresponding to channel 460.}\label{abb8}
\end{figure}

Further details concerning the properties of the coasting beam follow
from the measurement performed during machine studies where HERA was
filled with 10 proton bunches. The contribution of these protons
interacting with the target is shown in fig.\ref{abb8} in the time
interval corresponding to channels 60 to 100. The coasting beam was 
excited by a kicker
magnet in a narrow time slice around channel 460 every $21.12~ \mu s$,
i.e. once per full turn of a fill.
The measured distribution of the interaction rate (fig.\ref{abb8})
shows an exponential increase of the rate which starts at channel
$\sim 300$, peaks at channel $\sim 460$ and is followed by a sharp decrease.

The slope of the exponential corresponds to a characteristic ``lifetime''
of the protons of $\sim 0.5~\mu s$ which can be explained as follows.
The fraction of the coasting beam excited by the kicker produces a 
continuous beam of protons with large transverse emittance traveling around 
the storage ring with a slightly shorter circulation time than the 
synchronous protons since for coasting beam protons $\alpha_p > 0$.
Due to interactions of the protons with the target, the intensity of
the excited beam component decreases with time.

A proton interacts in the target after typically $10^5$ revolutions 
\cite{Ja}. From fig.\ref{abb8} one concludes that at this time it has 
advanced by $\sim 0.5~\mu s$ with respect to the synchronous protons if it
belongs to the coasting beam. Therefore after $n = 21.12\,\mu s/0.5\,\mu s
\cdot 10^5$ turns corresponding to a typical scale of 
\mbox{$t = 21.12\,\mu s \cdot n \approx 90\,s$}, the non--interacting 
excited protons in the coasting beam are smeared homogenously around the 
ring. This time has to be compared to the time needed by a proton close to 
the separatrix to advance by a full turn with respect to a centroid proton. 
The order of magnitude is given by

$$ t = \frac{L}{\Delta L} \cdot 21.12~\mu s = \frac{1}{\alpha_p \cdot
  \frac{\Delta E}{E}}~21.12~\mu s \approx 80~s$$

where $\frac{\Delta L}{L}$ is the relative deviation of the path length
for coasting protons, \mbox{$\alpha_p = 1.3~\cdot 10^{-3}$} is the momentum 
compaction factor of HERA and \mbox{$\left( \frac{\Delta E}{E} \right)_S 
\approx (2 \dots 3) \cdot 10^{-4}$} the energy deviation of typical protons  
close to the separatrix. Given the fact that the estimates are quite rough, 
the two time scales are in remarkable agreement.

\subsection{Impact of coasting beam on target operation}

\begin{figure}[ht!]
\begin{center}
\epsfig{file=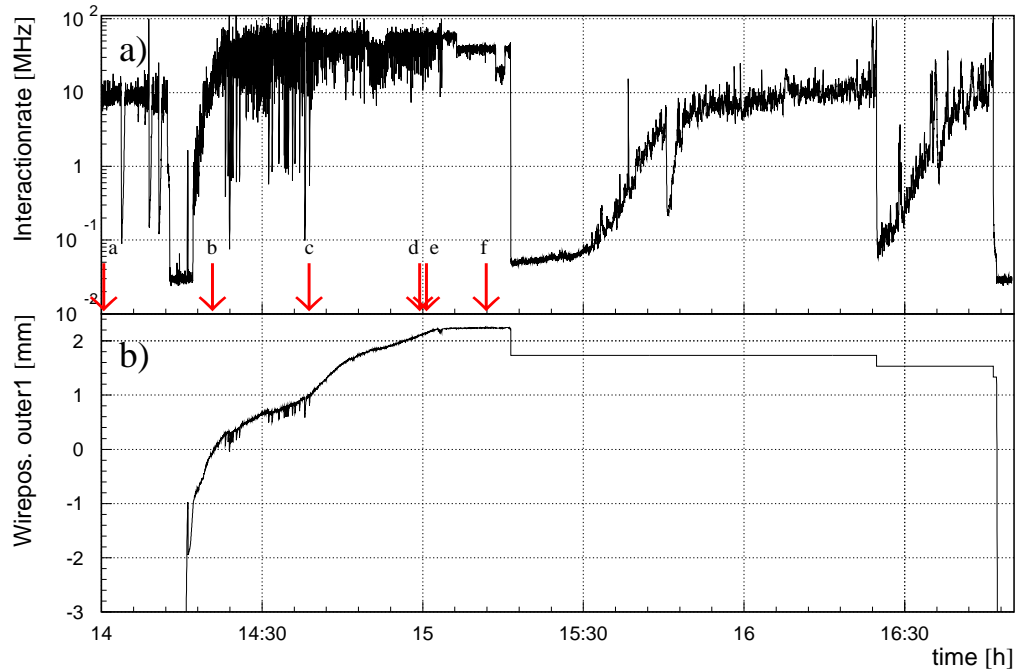, width=14cm}
\end{center}
\caption{a) Measured total rate; b) nominal outer wire position as
  function of time. For $t\le$ 14:12 the beam interacted with an
  inner target. For these data no precise vertex position information is 
  available, hence the absolute wire position with respect to the beam
  center is not known.}
\label{abb9}
\end{figure}

\begin{figure}[ht!]
\begin{center}
\epsfig{file=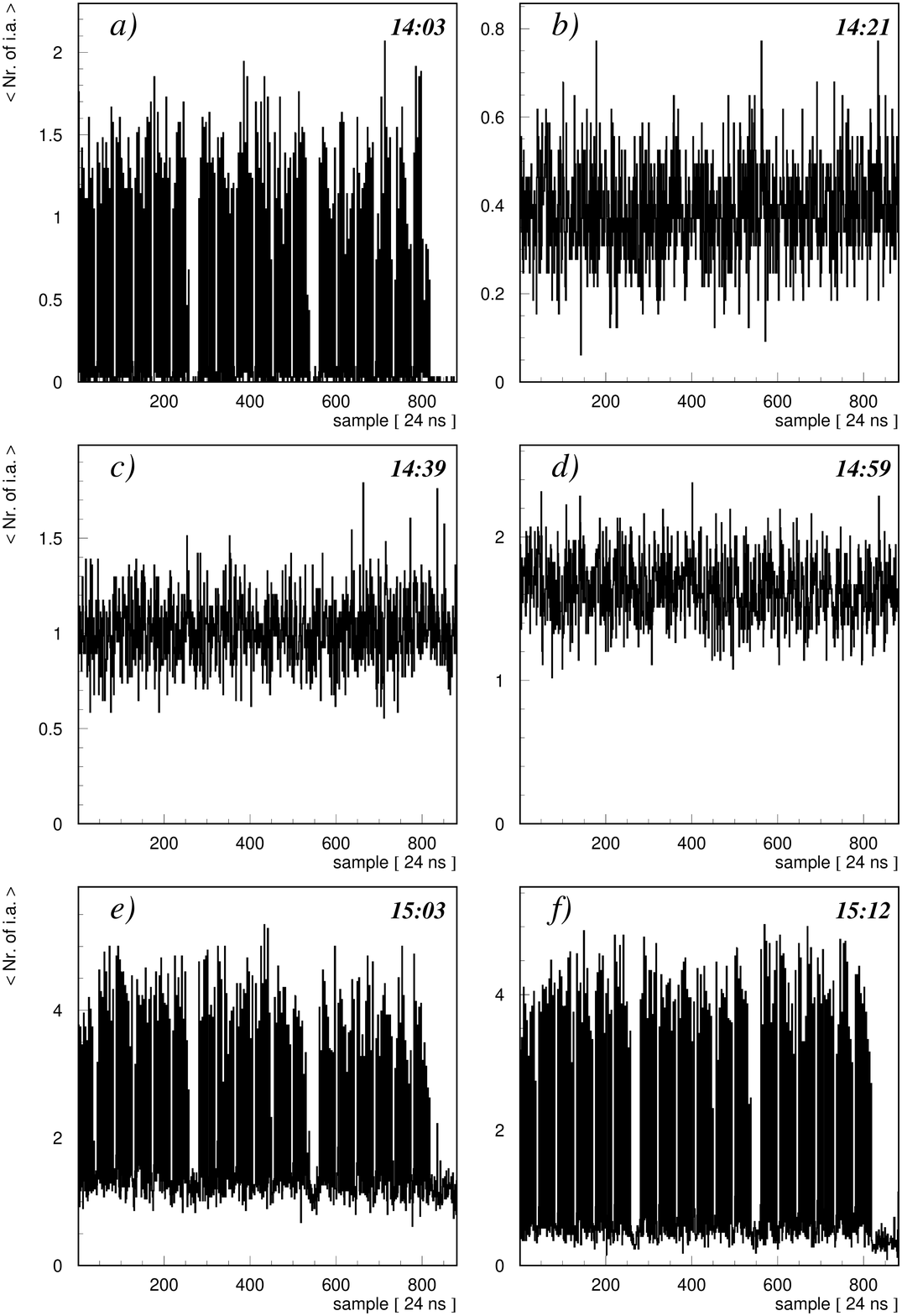, width=12.5cm}
\end{center}
\caption{Rates measured with a FADC at different times indicated by an
  arrow in fig.\ref{abb9}. The FADC spectra were collected at times 
  corresponding to the arrows in fig.\ref{abb9}.}\label{abb10}
\end{figure}

Measurements performed with an old fill are shown in
figs.\ref{abb9},~\ref{abb10}. Until t=14:12 the inner target was
operated with a constant interaction rate of $\approx$ 10~MHz. The
continuous background due to coasting beams is negligible
(fig.\ref{abb10}a), and the contribution of bunch correlated
interaction dominates. At the time 14:12, the inner target was
retracted and the outer target was inserted (fig.\ref{abb9}b). The
results shown in fig.\ref{abb10}b--d demonstrate that now nearly 100 \%
of the interaction rate measured under these conditions is due to
coasting beam protons. Initially, strong fluctuations of the
interaction rate are observed while the wire is gradually scraping
away the halo and moving towards the beam core (fig.\ref{abb9}a). At
$t$ = 15:00 the fluctuations disappear abruptly and the wire stops
moving towards the beam. At the same time, the coasting beam
background drops by a factor of two, as can be seen from data taken at
the times 14:59 (fig.\ref{abb10}d) and 15:03 (fig.\ref{abb10}e). With
further operation of the outer target wire at constant interaction
rate, the coasting beam background thereafter continues to decrease
slowly (fig.\ref{abb10}f). These effects can be interpreted by the
transition of the wire moving from the pure coasting beam halo into
the beam core region. In this sense the outer wire acts effectively
like a scraper for the coasting beam halo.

Similar to the measurements described in section 2.1 a steady increase
of the rate with time (fig.\ref{abb9}b) is observed, indicating the 
repopulation of the halo, if the outer wire is retracted by $0.5~mm$
(fig.\ref{abb9}a). The time structure of the proton bunches disappeares
completely and the interactions are again caused by the remaining coasting
beam component at large betatron amplitudes.

Note that the time dependence for the two rate measurements presented
above are similar and reproducible. They are indicative of the
inherent beam dynamics, leading to beam diffusion describable by the
Fokker--Planck equation \cite{Ma}. The sudden drop of the coasting beam
contribution observed at  $t =$ 15:00 in the measurement discussed
above deserves special attention and will be further analyzed by time
resolved measurements which are presently being prepared.


\section{Conclusions}

We have detected a continuous current of protons in the HERA machine
which produces nonbunched background at the HERA--B target.
Since the experiment is positioned at a location with a negative
dispersion, this observation can be attributed to protons with a smaller
energy than the synchronous ones. The measurement of the circulation time 
difference of protons for the coasting beam and synchronous protons and its 
quantitative agreement with estimates from linear beam optics support this
interpretation. We have shown that the coasting beams exist already for a 
virgin fill of the proton ring which is not disturbed in advance by proton 
interactions with the target. The interactions of bunched protons with the 
target increases the intensity of the coasting beam only marginally.

Moreover the protons of the coasting beam diffuse faster into the halo
region than the bunch correlated protons. A surprising observation is
the simultaneous sudden decrease of the rate fluctuations and the
coasting beam contribution at a characteristic wire--beam distance,
indicating a transition from the pure coasting beam halo into the
(mainly bunched) beam core.
With an improved setup we plan to study the effects with higher
time resolution. 

\newpage

\subsection*{Acknowledgement} 
This work was supported by the BMBF Bonn under contract number 05
7Do55P and 057Bu35I.
The observations presented in this paper and their understanding has been 
obtained in close and fruitful collaboration with the HERA machine
physicists. All further progress to understand and overcome the
problems related with coasting beam at HERA requires still a close
cooperation and the support of the HERA machine group.

We like to express our cordial thanks to the HERA crew for the friendly 
collaboration, their support and their assistance to discuss our
observations and to carry out dedicated machine studies to investigate 
and improve the situation. 

It is impossible to name all of them, special thanks we owe to 
Jim Ellis, B. Holzer, Jens Kluthe, Helmut Mais, Chr. Montag, 
Mark Lomperski,  
 Tanaji Sen,  F. Willeke and  Mari Paz Zorzano.

\end{document}